\documentclass[10pt]{article}
\usepackage{graphicx}

\def\Title#1{\begin{center} {\Large #1 } \end{center}}
\def\Author#1{\begin{center}{ \sc #1} \end{center}}
\def\Address#1{\begin{center}{ \it #1} \end{center}}

\newcommand\pubblock{\rightline{\begin{tabular}{l} Proceedings of the Second Annual LHCP\\ \pubnumber\\
         \pubdate  \end{tabular}}}

\newenvironment{Abstract}{\begin{quotation} \begin{center} 
             \large ABSTRACT \end{center}\bigskip 
      \begin{center}\begin{large}}{\end{large}\end{center} \end{quotation}}

\newenvironment{Presented}{\begin{quotation} \begin{center} 
             PRESENTED AT\end{center}\bigskip 
      \begin{center}\begin{large}}{\end{large}\end{center} \end{quotation}}

\def\beq{\begin{equation}}
\def\eeq#1{\label{#1}\end{equation}}
\def\eeqn{\end{equation}}

\def\beqa{\begin{eqnarray}}
\def\eeqa#1{\label{#1}\end{eqnarray}}
\def\eeqan{\end{eqnarray}}

\let\bar=\overbar

\def\Dslash{\not{\hbox{\kern-4pt $D$}}}
\def\dslash{\not{\hbox{\kern-2pt $\del$}}}

\def\msb{{\bar{\ssstyle M \kern -1pt S}}}

\usepackage{color}

\newcommand\pubnumber{ CMS CR-2014/217 }

\newcommand\pubdate{\today}

\def\affiliation{
On behalf of the CMS Experiment \\
\vspace{0.3cm}
Dipartimento Interateneo di Fisica, \\
Universit\'{a} degli Studi di Bari ``Aldo Moro'' and INFN-Sezione di Bari, Italy}

\begin{document}

\large
\begin{titlepage}
\pubblock

\vspace{2cm}

\vfill
\Title{QUARKONIUM PRODUCTION AND POLARIZATION \\IN PP COLLISIONS WITH THE CMS DETECTOR}
\vfill

\Author{ALEXIS POMPILI}
\Address{\affiliation}
\vfill
\begin{Abstract}

The studies of heavy quarkonium inclusive production and polarization at LHC are becoming crucial to solve the puzzle of hadron formation.

The results by CMS and the other LHC experiments are compactly presented 

for the five S-wave states $J/\psi$, $\psi(2S)$ and $\Upsilon(nS)$ ($n=1,2,3$) 

and briefly discussed especially in comparison to the theoretical predictions 

provided by Non Relativistic QCD.

\end{Abstract}
\vfill

\begin{Presented}
The Second Annual Conference \\
on Large Hadron Collider Physics \\
Columbia University, New York, U.S.A \\ 
June 2-7, 2014
\end{Presented}
\vfill
\end{titlepage}
\def\thefootnote{\fnsymbol{footnote}}
\setcounter{footnote}{0}
%

\normalsize 


\section{Introduction}

Quarkonia are bound states of an heavy-quark and an heavy-antiquark ($c\bar{c}$, $b\bar{b}$) and exist in ``families'' of several colorless states (neutral mesons).
Quarkonium spectra and decays are well understood below open charm and open beauty thresholds.
Quarkonium production is still an active field of research; production rates at the LHC are rather high and LHC can be considered a ``quarkonium factory''.
Quarkonium production occurs through two mechanisms: a) \textit{prompt} production, \textit{direct} or \textit{feed-down} from higher quarkonium states, 
b) \textit{non-prompt} production, from B decays (for charmonia only).
The study of quarkonium prompt production is suited to understand how quarks combine into a bound state (the hadron). This mechanism is still not easy to understand, 
likely because it is a part of the non-perturbative QCD sector. 

Properties of QCD can be probed by LHC experiments, in different new kinematic regions, through several quarkonium production measurements including production cross sections 
(discussed in section 2) and polarizations (discussed in section 3). Many extensive and detailed reviews about these topics are available 
and, among them, \cite{fac14:faccioli2014}\cite{bod:bodwin2013}\cite{fac10:faccioli2010}\cite{brat:braaten2014} are certainly useful to enter this field.

\section{Quarkonium Production}

In the $Q\bar{Q}$ center-of-mass frame quarks can be considered not relativistic;
being $v$ the velocity of the heavy-quark in this frame, $\nu = v/c \ll 1 $ is a better approximation for $b\bar{b}$ than $c\bar{c}$.
The non-relativistic QCD (NRQCD) is an effective field theory that treats heavy quarkonia as non-relativistic systems \cite{nrqcd}.
The heavy-quark mass is $m_{Q} \gg \Lambda_{QCD}$ which provides a natural boundary between short- and long-distance QCD. 
This allows the inclusive quarkonium production to be factorized in two distinct steps.
In the first step the $Q\bar{Q}$ pair should be produced in the regime of perturbative QCD, whereas in the second step a bound state driven by non-perturbative QCD should be formed.
In this NRQCD factorization conjecture the inclusive cross section for producing the quarkonium state $H$ with enough large momentum transfer $p_{T}$ is the sum of short-distance
coefficients (SDCs, $\sigma_{n}$) multiplied by long-distance matrix elements (LDMEs, $P_{n}$)

\[ \sigma(A+B \rightarrow H+X) = \sum_{n} \sigma_{n} \cdot P_{n} = \sum_{n} \sigma (A+B \rightarrow [Q\bar{Q}]_{n} + X) \cdot P([Q\bar{Q}]_{n} \rightarrow H) \] 
where the sum is over the states with $n = \: ^{2S+1}L_{J}^{C}$ since the $Q\bar{Q}$ pair can be, at short distances, produced in a state $n$ with definite spin $S$, 
angular momentum $L$, total angular momentum $J$ and color $C=1,...,8$.

The SDCs are inclusive perturbative QCD cross sections of partonic processes to form the $Q\bar{Q}$ pair in state $n$ (convoluted with probability density functions); 
they are process- and kinematics-dependent functions calculated perturbatively as expansions in $\alpha_{s}$. 
On the other hand the LDMEs represent the probability of the $Q\bar{Q}$ pair in state $n$ to evolve into the quarkonium final state $H$;
they are expected to be universal constants, i.e. the same in $pp$ collisions, $e^{+}e^{-}$ collisions and in general for any collision process and independent of kinematics 
variables (such as $p_{T}$ and the rapidity $y$). The LDMEs are determined by fits to experimental data and their relative importance should depend on $\nu$-scaling rules.

Theoretical predictions are organized as double expansions in $\alpha_{s}$ and $\nu$.
In the current NRQCD phenomenology the truncation of $\nu$-expansion for spin-triplet S-wave states ($J/\psi$, $\psi(2S)$; $Y(1S)$, $Y(2S)$ and $Y(3S)$) includes four terms:
the Color Singlet (CS) term ($^3 S_{1}^{[1]}$) and three Color Octet (CO) terms ($^1 S_{0}^{[8]}$, $^3 S_{1}^{[8]}$ and $^3 P_{J=0,1,2}^{[8]}$), 
all of relative order $O(\nu^{4})$ with respect to the CS term.
The CS term is characterized by a suppression of powers of $\alpha_{s}$ thus making important the CO channels despite of their suppression by powers of $\nu$.
The old color singlet model \cite{csm} assumed that the initial $Q\bar{Q}$ pair and the final quarkonium $H$ have the same quantum numbers, whereas NRQCD predicts the existence
of intermediate color octet states that subsequently evolve into color singlet quarkonia by non-perturbative emission of soft gluons.

\begin{figure}[htb]
\centering
\includegraphics[height=7.5cm]{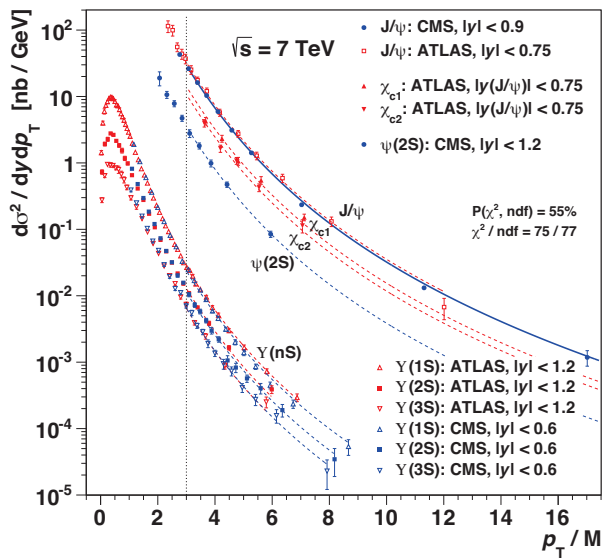}
\caption{This compilation \cite{fac14:faccioli2014} summarizes results from CMS \cite{cmsprod:charm}\cite{cmsprod:bottom} and ATLAS \cite{atlas:prod1}\cite{atlas:prod2}\cite{atlas:prod3}. 
$p_{T}$ is mass rescaled to equalize the kinematics' effects of different average parton momenta and phase spaces. 
The solid curve represents the fit to CMS $J/\psi$ data whereas the dashed curves are replicas with adjusted normalizations.}  
\label{fig:prodreview}
\end{figure}

NRQCD provides \cite{buten:kniel}\cite{gong:etal} good consistency in the fit of the differential production cross sections as a function of $p_{T}$ 
when including CO terms and next-to-leading order (NLO) corrections in $\alpha_{s}$;
the fit function consists in a superposition of CS and CO SDCs and the CS contribution is fixed while the CO terms 
characterized by fixed shape but floating normalizations (represented by the LDMEs).

Fig.\ref{fig:prodreview} shows a compilation \cite{fac14:faccioli2014} of mid-rapidity double differential cross sections for the production of 7 different quarkonia 
as a function of $p_{T}/M$. All shapes are well described by a single empirical power-law function for $p_{T}/M > 3$,
a cut justified by the fact that NRQCD calculations are not expected to reproduce the measurements at low $p_{T}$ values.
This $p_{T}/M$ scaling behaviour, common to five S-wave states and two P-wave states characterized by different feed-down contamination levels, suggests a simple composition
of processes dominated by one single mechanism; in particular one single CO term should dominate their production and this could be $^1S_{0}^{[8]}$ if the NRQCD fit would 
start at $p_{T}/M > 3$ as argumented in \cite{fac14:faccioli2014}.

This scaling behaviour needs to be confirmed with more accurate data up to higher $p_{T}$ values since the fits also suggest that $^3 S_{1}^{[8]}$ could become dominant at higher momenta. 
During the preparation of this proceeding CMS made available updated production measurements, 
based on the full 2011 dataset \cite{bph-14-001}, covering a broader $p_{T}$ range extending from $15$ to $120$~GeV for the $J/\psi$ and up to $100$~GeV for the $\psi(2S)$.

\section{Quarkonium Polarization}

The polarization is sensitive to the hadroproduction mechanism and therefore is important for the theoretical understanding.
The polarization of a vector meson decaying into a leptons pair is reflected in the leptons' angular distributions, 
specified in terms of the spherical angles $\theta$ and $\phi$ of the momentum vector of the positively charged lepton in the meson rest frame.
A polarization frame must be chosen to define these two angles: $\theta$ is the polar angle with respect to the spin-quantization $z$-axis,
whereas $\phi$ is the azimuthal angle with respect to the $x$-axis that lies, together with the $z$-azis, in the collision plane (defined by the momentum vectors
of the colliding hadrons boosted in the meson rest frame). However there is not an unique choice of the polarization $z$-axis in the collision plane and, correspondingly,
at least three conventional reference frames can be introduced \cite{fac10:faccioli2010}: 
a) center-of-mass helicity frame (HX), 
b) Collins-Soper frame (CS), and 
c) perpendicular helicity frame (PX).

The most general 2D angular distribution for the dileptons from the decay of a vector meson is specified by three polarization parameters $\lambda_{\theta}$, 
$\lambda_{\phi}$, $\lambda_{\theta\phi}$ and can be expressed, apart from a normalization factor, as follows:

\[ W \equiv \frac{d^{2}N}{d(cos\theta)d\phi} \propto \frac{1}{(3+\lambda_{\theta})} 
( 1 + \lambda_{\theta} cos^{2}\theta  + \lambda_{\phi} sin^{2}\theta cos2\phi + \lambda_{\theta \phi} sin2\theta cos\phi )\]

\noindent Two extreme angular decay distributions obtainable by ($\lambda_{\theta} = \pm 1$, $\lambda_{\phi} = 0$, $\lambda_{\theta\phi} =0$) represent
transverse ($J_{z}=\pm 1$) and longitudinal ($J_{z}=0$) polarization, respectively. Thus $\lambda_{\theta}$ measures the degree of polarization with respect to 
the spin-quantization axis. Theoretically \cite{brat:braaten2014} this parameter can be expressed as a function of the cross sections for the two transverse states ($\sigma_{T}$) and the
longitudinal state ($\sigma_{L}$) as follows: $\lambda_{\theta} = (\sigma_{T}-2\sigma_{L}$)/($\sigma_{T}+2\sigma_{L}$). 
Generally $\lambda_{\theta}$, $\lambda_{\phi}$ and $\lambda_{\theta\phi}$ can be expressed as a function of components of the $p\bar{p} \rightarrow H + X (H\equiv J/\psi,\psi(2S),...)$ differential 
cross sections given in terms of PDFs and partonic spin density matrix elements.

Each NRQCD term is characterized by a specific polarization: at NLO, $^3 S_{1}^{[1]}$ is longitudinal, $^1 S_{0}^{[8]}$ isotropic, $^3 S_{1}^{[8]}$ transverse at high $p_{T}$ values 
and $^3 P_{J}^{[8]}$ hyper-transverse at high $p_{T}$.

As explained in \cite{fac10:faccioli2010} the observed polarization depends on the polarization frame and the polarization can be fully determined when either 
a) both the polar and azimuthal components of angular distribution $W$ are known ($\lambda_{\theta}$ and $\lambda_{\phi}$) or
b) a single polarization parameter is measured in at least two complementary polarization frames (for instance $\lambda_{\theta}$ in the HX and CS frames).
On the other hand, being the shape of the angular distribution frame-independent, it can be characterized by a frame-invariant combination of the parameters 
such as $\tilde{\lambda} = (\lambda_{\theta} + 3\lambda_{\phi})$ / $(1-\lambda_{\phi})$ \cite{facprl:faccioli}.

The three polarization parameters and $\tilde{\lambda}$ have been measured by CMS \cite{cmspol:charm}\cite{cmspol:bottom} in three different frames (HX,CS,PX) for the 5 S-wave states as a 
function of transverse momentum and rapidity. No evidence of strong longitudinal or transverse polarizations has been observed, independently of the level of feed-down contributions
(with unknown polarizations) characterizing differently these states. This result holds for any parameter in any frame; moreover there is a good agreement among the $\tilde{\lambda}$
values in the three polarizations frames thus showing that the results are consistent. 

\begin{figure}[hb!]
\centering
\begin{tabular}{cc}
\includegraphics[height=5.89cm]{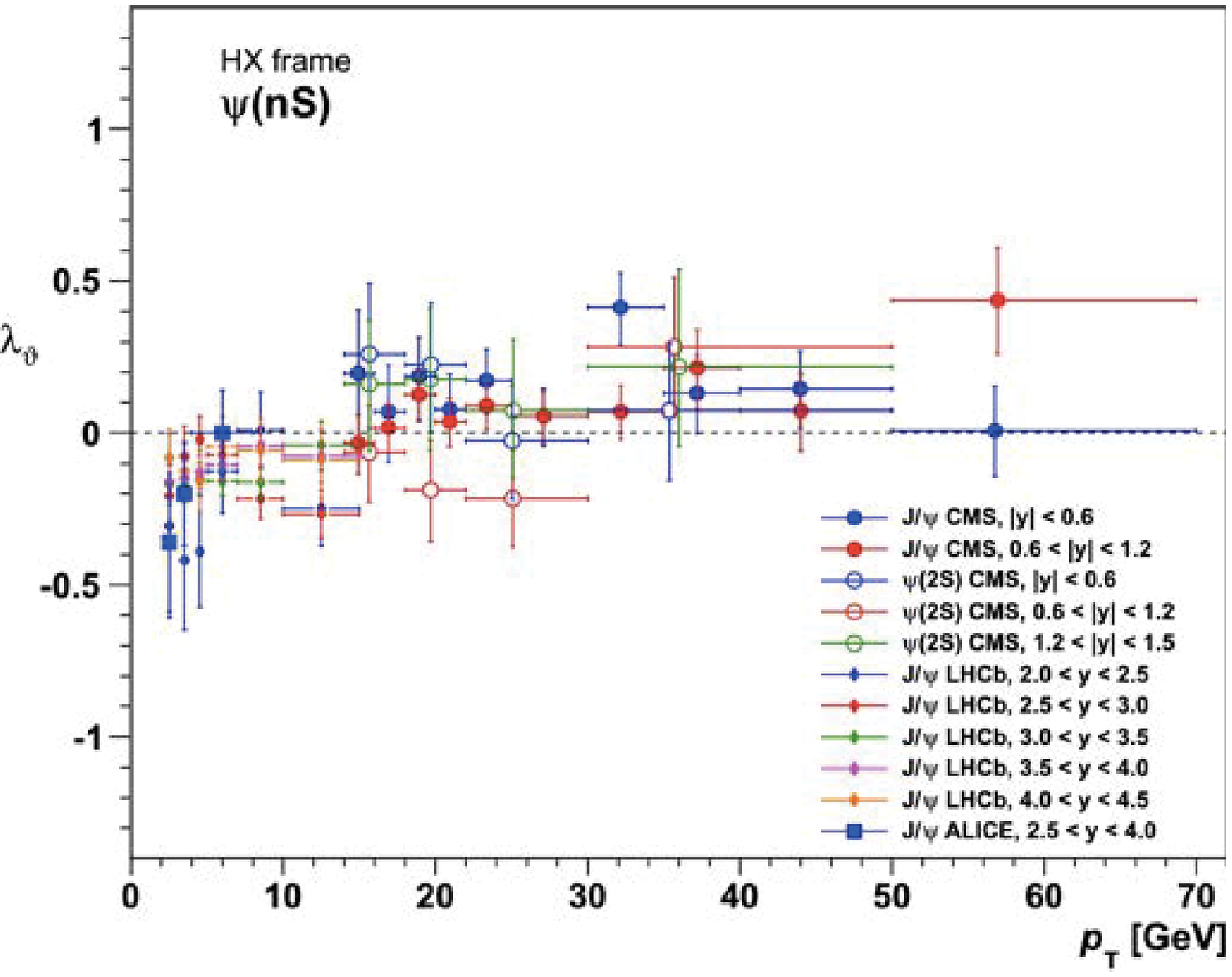} &
\includegraphics[height=6.1cm]{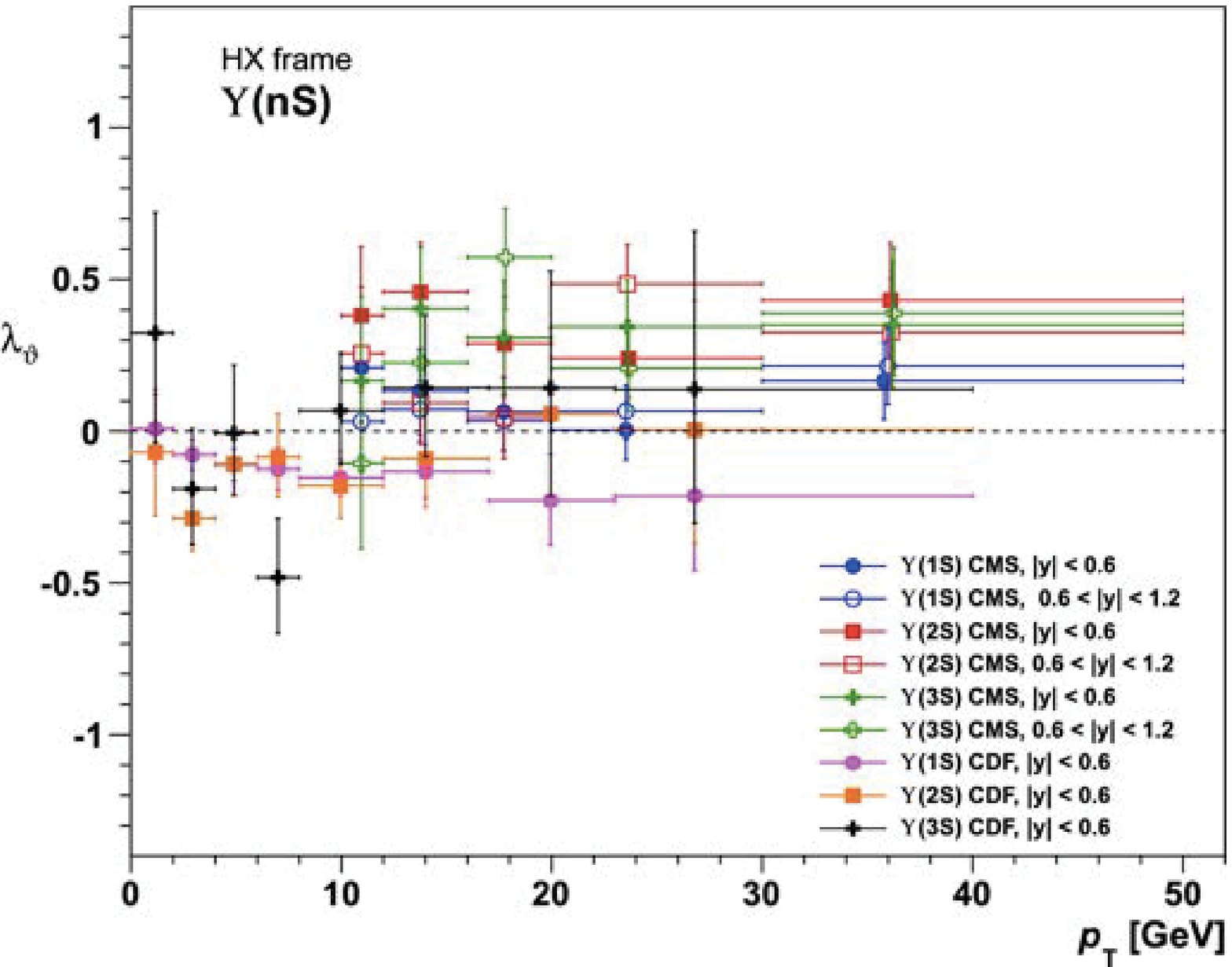} \\
\end{tabular}
\caption{This compilation \cite{fac14:faccioli2014} summarizes polarization results from CMS \cite{cmspol:charm}\cite{cmspol:bottom}, 
LHCb\cite{lhcbpol:charm}, Alice\cite{alicepol:charm} and CDF\cite{cdfpol:bottom}: $\lambda_{\theta}$ is measured the HX frame, as a function of $p_{T}$, in different rapidity intervals.}
\label{fig:polreview}
\end{figure}

Fig.\ref{fig:polreview} presents a compilation \cite{fac14:faccioli2014} of $\lambda_{\theta}$ measurements in the HX frame at LHC, showing that all LHC results are compatible with each other.
The polarizations cluster around the unpolarized limit with 
a) no significant dependencies on $p_{T}$ or $y$, 
b) no strong changes among states
differently affected by P-wave feed-down and the $\psi(2S)$ not affected at all, 
c) no evident differences between $c\bar{c}$ and $b\bar{b}$ states.
This again suggests \cite{fac14:faccioli2014} that all quarkonia are predominantly produced by a single mechanism and that, given the unpolarized result, 
the dominant contribution must come from $^1S_{0}^{[8]}$. New $p_{T}$-extended polarization measurements would allow to test if polarization turns to transverse accordingly to a production dominated at higher momenta by 
the $^3 S_{1}^{[8]}$ contribution.

\vspace{0.4cm}

\newpage

\begin{figure}[ht!]
\centering
\begin{tabular}{cc}
\includegraphics[height=5.8cm]{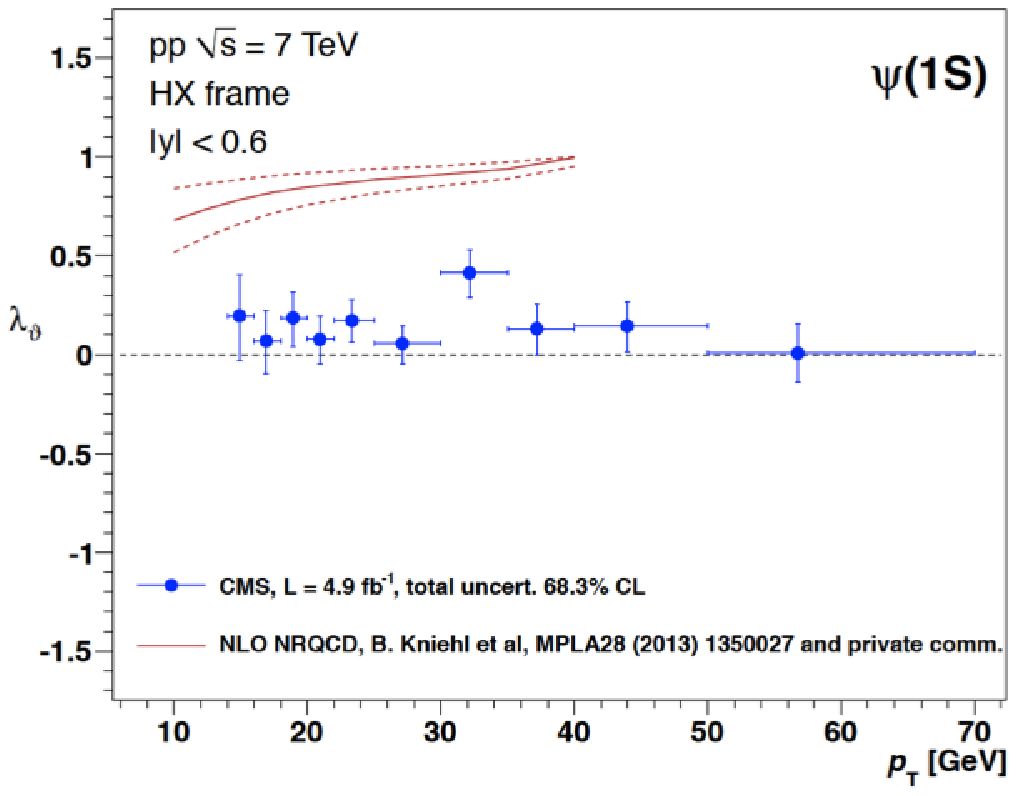} &
\includegraphics[height=5.8cm]{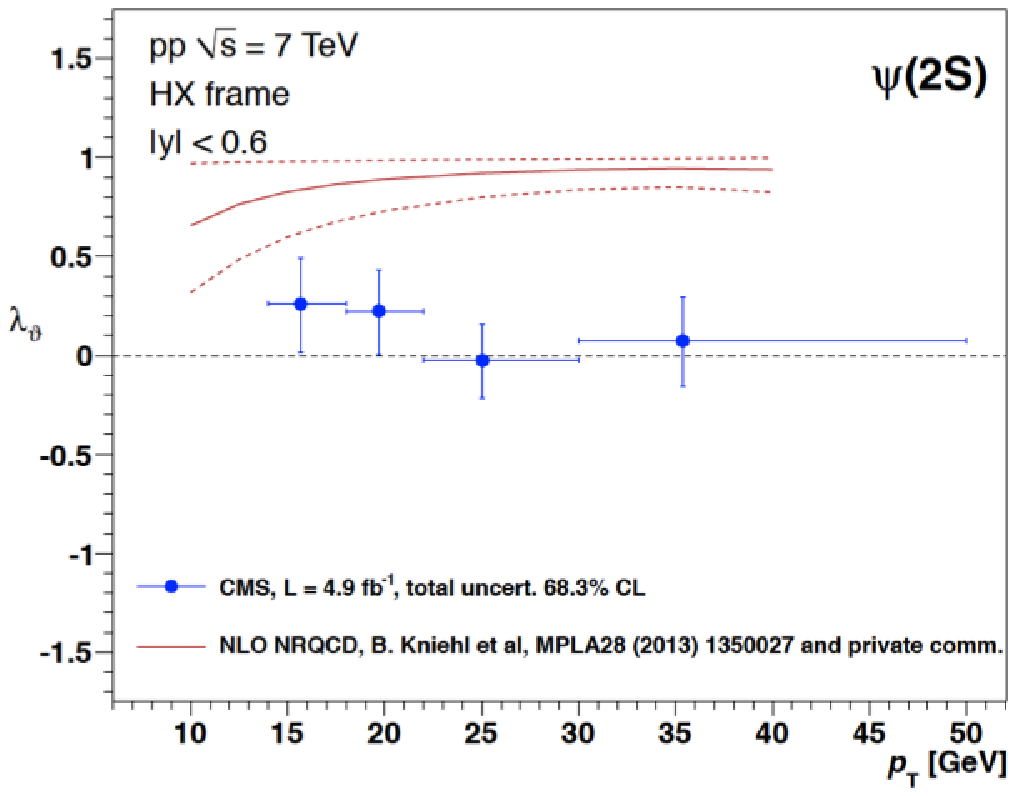} \\
\end{tabular}
\caption{$\lambda_{\theta}$ parameter for $J/\psi$ (left) and $\psi(2S)$ (right) states measured by CMS \cite{cmspol:charm} in the HX frame,
as a function of dimuon $p_{T}$ for $|y|<0.6$, compared to NLO NRQCD predictions \cite{buten:kniel}.}
\label{fig:cNRQCDcomp}
\end{figure}

\vspace{0.4cm}

\begin{figure}[hb!]
\centering
\begin{tabular}{ccc}
\includegraphics[height=4cm]{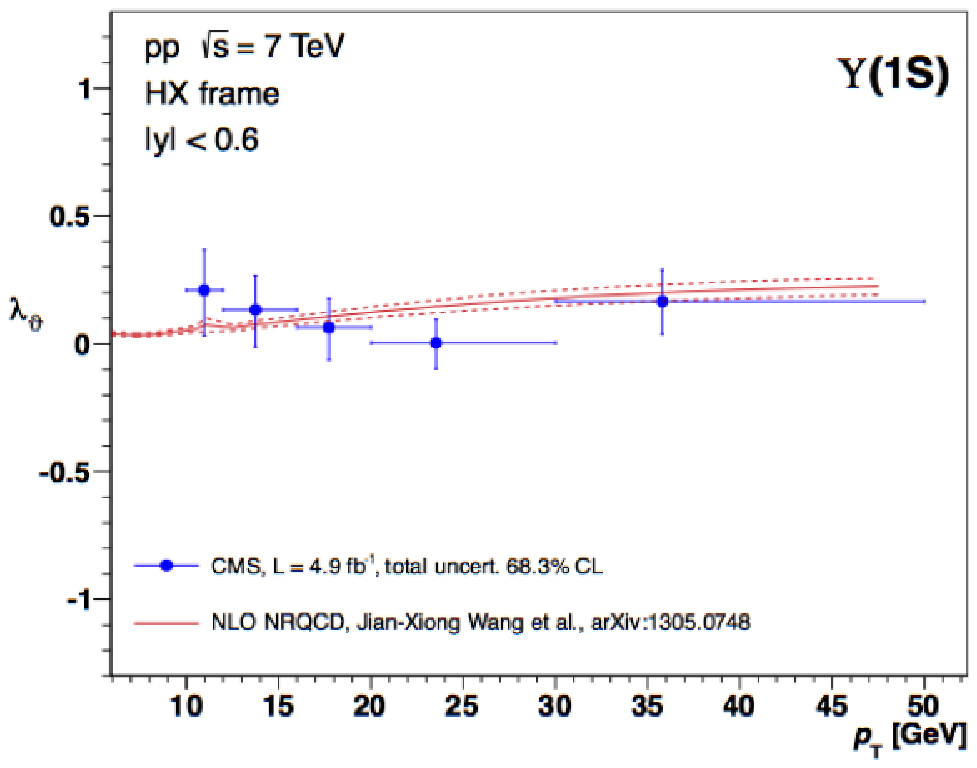} &
\includegraphics[height=4cm]{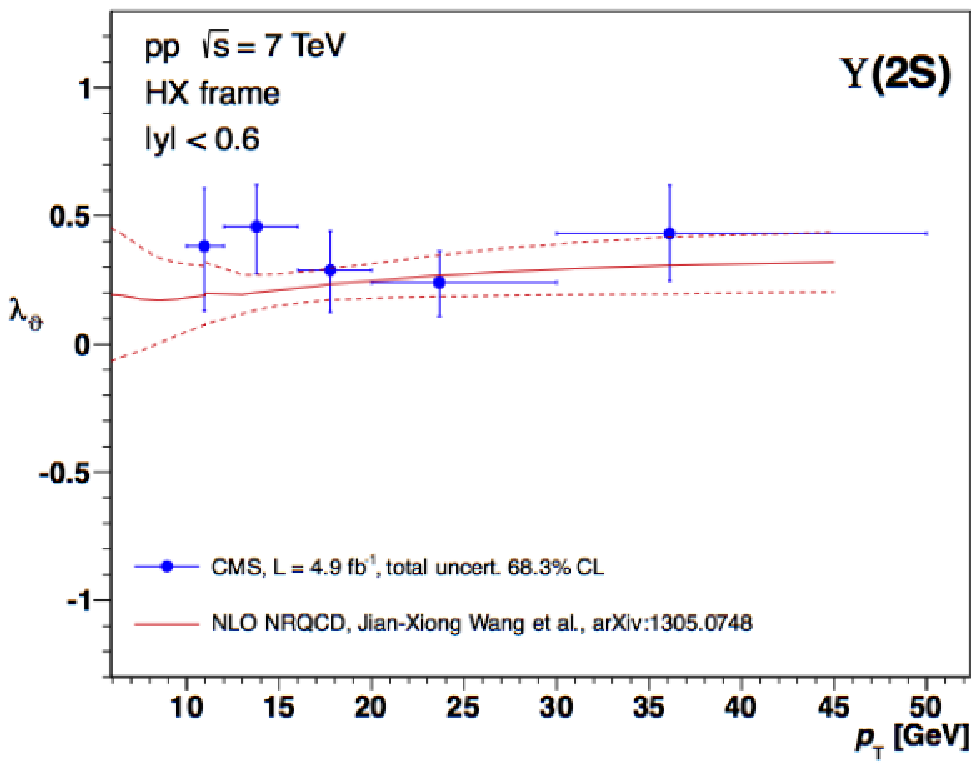} &
\includegraphics[height=4cm]{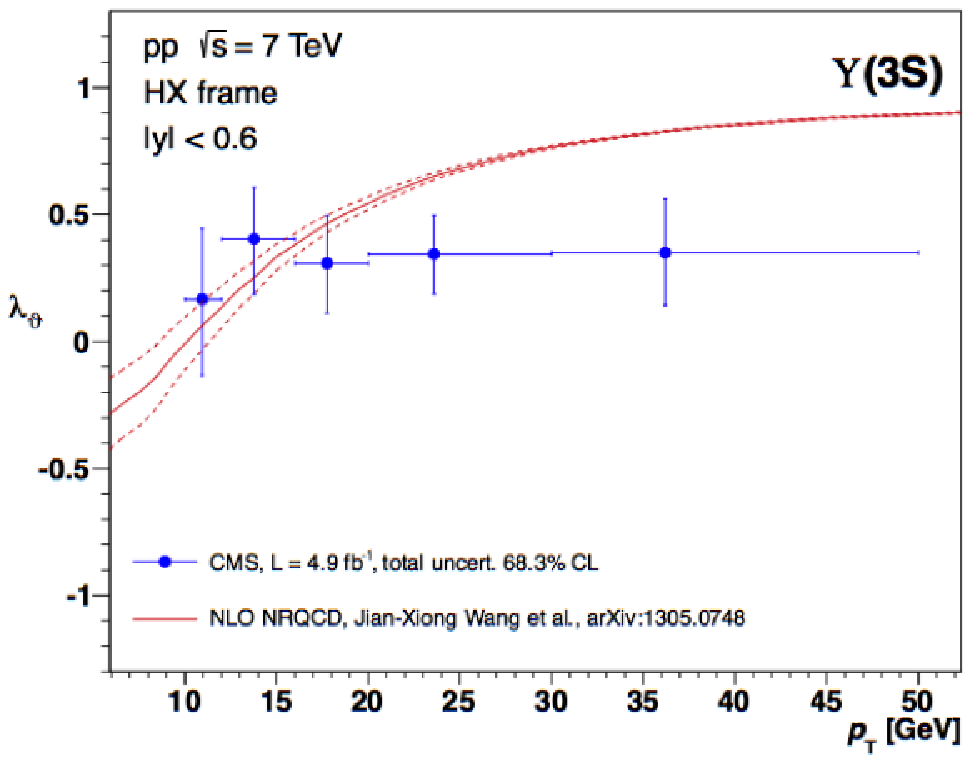} \\
\end{tabular}
\caption{$\lambda_{\theta}$ parameter for $\Upsilon(1S)$ (left), $\Upsilon(2S)$ (center) and $\Upsilon(3S)$ (right) states measured by CMS \cite{cmspol:bottom} in the HX frame,
as a function of dimuon $p_{T}$ for $|y|<0.6$, compared to NLO NRQCD predictions \cite{gong:etal}.}
\label{fig:bNRQCDcomp}
\end{figure}

Figg.\ref{fig:cNRQCDcomp} and \ref{fig:bNRQCDcomp} show the $\lambda_{\theta}$ parameter measured in the HX frame by CMS \cite{cmspol:charm}\cite{cmspol:bottom} in
comparison to NLO NRQCD predictions \cite{buten:kniel}\cite{gong:etal}, for the 5 S-wave states. 

In Fig.\ref{fig:cNRQCDcomp} the theoretical calculations use global fit of CO LDMEs to photo- as well as hadro-production data (polarization results are not considered);
their predictions only consider direct production and therefore the comparison makes sense only for $\psi(2S)$ data since it should not suffer from feed-down contributions
with unknown polarizations. The disagreement between theoretical and experimental results is well evident. 

In the theoretical calculations of Fig.\ref{fig:bNRQCDcomp} CO LDMEs are left as free parameters in the fit to hadro-production data;
these calculations include the feed-down contributions to $\Upsilon(1S)$ and $\Upsilon(2S)$ and the extra adjustable fit parameters (LDMEs of $\chi_{bJ}(nP)$ states) help
reaching compatibility with data. The $\Upsilon(3S)$ state instead is assumed to be exclusively directly produced, thus allowing less freedom in the fit and related disagreement between
theoretical predictions and experimental measurements. During the preparation of this proceeding LHCb has revealed the observation that about $50\%$ of $\Upsilon(3S)$ mesons
are not produced directly but they originate from the radiative decay of $\chi_{b}(3P)$ mesons \cite{lhcb:chib3p}. Thus the assumption of a negligible feed-down contribution 
to the prompt production of $\Upsilon(3S)$ needs to be revisited.

\vspace{0.5cm}

\section{Conclusions}

Cross section measurements at LHC are dominated by color octet production. In the hadron formation 
the $Q\bar{Q}$ bound states seem to be preferably formed by two heavy-quarks of different colors
and smaller relative angular momentum and spin and only later evolve into the physical color singlet quarkonia.
Polarization measurements show no relevant longitudinal or transverse polarization for S-wave states,
in disagreement with NLO NRQCD predictions. Discrepancy between theory and experiment with respect to polarization
deserves further investigation both theoretically and experimentally. 

Theory-data comparisons should be reconsidered including polarization data in global NRQCD analyses of production.
In the experimental field it is essential to perform measurements with smaller uncertainties and extend $p_{T}$ reach thus testing the NRQCD validity domain.
It would be useful for future measurements to separate the feed-down contributions from direct production.
Challenging measurements of production and polarization for the $\chi_{cJ}$ and $\chi_{bJ}$ P-wave states would provide valuable additional tests.



\begin{thebibliography}{99}


\bibitem{fac14:faccioli2014} 
P.~Faccioli {\it et al.}, Phys.\ Lett.\ B {\bf 736}, 98 (2014).

\bibitem{bod:bodwin2013}
G.T.~Bodwin {\it et al.}, arXiv:1307.7425, Conference C13-07-29.2 (Snowmass 2013)

\bibitem{fac10:faccioli2010} 
P.~Faccioli {\it et al.}, Eur. Phys. J. C {\bf 69} (2010) 657

\bibitem{brat:braaten2014}
E.~Braaten and J.~Russ, arXiv:1401.7352 (for Vol. 64 of the Annual Review of Nucl. and Part. Science) 

\bibitem{nrqcd}
G.T.~Bodwin {\it et al}, Phys. Rev. D {\bf 51} (1995) 1125

\bibitem{csm}
E.L.~Berger and D.L.~Jones, Phys. Rev. D {\bf 23} (1981) 1521; \\
R.~Baier and R.~Ruckl, Phys.\ Lett.\ B {\bf 102}, 364 (1981).

\bibitem{buten:kniel}
M.~Butenschoen and B.A.~Kniehl, Phys. Rev. Lett. {\bf 108} (2012) 172002, later in Mod. Phys. Lett. A {\bf 28} (2013) 1350027, and private communication
\bibitem{gong:etal}
B.~Gong {\it et al.}, Phys. Rev. Lett. {\bf 112} (2014) 032001 

\bibitem{cmsprod:charm} 
CMS Collaboration, JHEP {\bf 02} (2012) 011

\bibitem{cmsprod:bottom} 
CMS Collaboration, Phys.\ Lett.\ B {\bf 727}, 101 (2013)

\bibitem{atlas:prod1} 
ATLAS Collaboration, JHEP {\bf 07} (2014) 154

\bibitem{atlas:prod2} 
ATLAS Collaboration, Nucl. Phys. B {\bf 850} (2011) 387

\bibitem{atlas:prod3} 
ATLAS Collaboration, Phys. Rev. D {\bf 87} (2013) 052004

\bibitem{bph-14-001}
https://twiki.cern.ch/twiki/bin/view/CMSPublic/PhysicsResultsBPH14001/


\bibitem{facprl:faccioli}
P.~Faccioli {\it et al.}, Phys. Rev. Lett. {\bf 105} (2010) 061601

\bibitem{cmspol:charm}
CMS Collaboration, Phys. Lett. B {\bf 727} (2013) 381

\bibitem{cmspol:bottom}
CMS Collaboration, Phys. Rev. Lett. {\bf 110} (2013) 081802

\bibitem{lhcbpol:charm}
LHCb Collaboration, Eur. Phys. J. C {\bf 73} (2013) 41

\bibitem{alicepol:charm}
Alice Collaboration, Phys. Rev. Lett. {\bf 108} (2012) 082001

\bibitem{cdfpol:bottom}
CDF Collaboration, Phys. Rev. Lett. {\bf 108} (2012) 151802

\bibitem{lhcb:chib3p}
LHCb Collaboration, arXiv:1407.7734 (2014)


\end{thebibliography}
\end{document}